\documentclass[aps,prl,twocolumn,reprint,nofootinbib,superscriptaddress,showkeys]{revtex4-1}

\usepackage{color}
\usepackage{graphicx}
\usepackage{dcolumn}
\usepackage{bm}
\usepackage{commath}
\usepackage{amsmath}
\usepackage{upgreek}
\usepackage{siunitx}
\usepackage{physics}
\usepackage[mathlines]{lineno}
\usepackage[english]{babel}
\usepackage[latin1]{inputenc}
\usepackage[bottom=2cm,top=2cm, left=1.5cm, right=1.5cm]{geometry}
\usepackage{booktabs}
\usepackage[unicode=true, bookmarks=true, bookmarksnumbered=false, bookmarksopen=false, breaklinks=false, pdfborder={0 0 1}, backref=false, colorlinks=false]{hyperref}
\usepackage[normalem]{ulem}
\usepackage{enumerate}
\usepackage{color}
\usepackage{dcolumn}
\usepackage{bm}
\usepackage{mathptmx}
\usepackage{pifont}
\usepackage{cancel}
\usepackage{amssymb}
\usepackage{lineno}
\usepackage{xcolor}

\hypersetup{pdftitle={Critical slowing down of fermions near a magnetic quantum phase transition}, pdfauthor={Chia-Jung Yang, Kristin Kliemt, Cornelius Krellner, Johann Kroha, Manfred Fiebig, and Shovon Pal}}

\begin{document}

\title{Critical slowing down of fermions near a magnetic quantum phase transition}

\author{Chia-Jung Yang}
\affiliation{Department of Materials, ETH Zurich, 8093 Zurich, Switzerland}

\author{Kristin Kliemt}
\affiliation{Physikalisches Institut, Goethe-Universit\"{a}t Frankfurt, 60438 Frankfurt, Germany}

\author{Cornelius Krellner}
\affiliation{Physikalisches Institut, Goethe-Universit\"{a}t Frankfurt, 60438 Frankfurt, Germany}

\author{Johann Kroha}
\affiliation{Physikalisches Institut and Bethe Center for Theoretical Physics, Universit\"{a}t Bonn, 53115 Bonn, Germany}

\author{Manfred Fiebig}
\affiliation{Department of Materials, ETH Zurich, 8093 Zurich, Switzerland}

\author{Shovon Pal}
\affiliation{Department of Materials, ETH Zurich, 8093 Zurich, Switzerland}
\affiliation{School of Physical Sciences, National Institute of Science Education and Research, HBNI, Jatni, 752 050 Odisha, India}

\date{\today}

\begin{abstract}
A universal phenomenon in phase transitions is critical slowing down (CSD) --- systems, after an initial perturbation, take an exceptionally long time to return to equilibrium. It is universally observed in the dynamics of bosonic excitations, like order-parameter collective modes, but it is not generally expected to occur for fermionic excitations because of the half-integer nature of the fermionic spin. Direct observation of CSD in fermionic excitations or quasiparticles would therefore be of fundamental significance. Here, we observe fermionic CSD in the heavy-fermion (HF) compound YbRh$_2$Si$_2$ by terahertz time-domain spectroscopy. HFs are compound objects with a strongly enhanced effective mass, composed of itinerant and localized electronic states. We see that near the quantum phase transition in YbRh$_2$Si$_2$ the build-up of spectral weight of the HFs towards the Kondo temperature $T_K\approx 25$\,K is followed by a logarithmic rise of the quasiparticle excitation rate on the heavy-Fermi-liquid side of the quantum phase transition below $10$~K. A critical two-band HF liquid theory shows that this is indicative of fermionic CSD. This CSD is a clear indication that the HF quasiparticles experience a breakdown near the quantum phase transition, and the critical exponent of this breakdown introduces a classification of fermionic quantum phase transitions analogous to thermodynamic phase transitions --- solution to a long-standing problem.
\end{abstract}

\maketitle

\begin{figure}
\centering
\includegraphics[width=0.9\columnwidth]{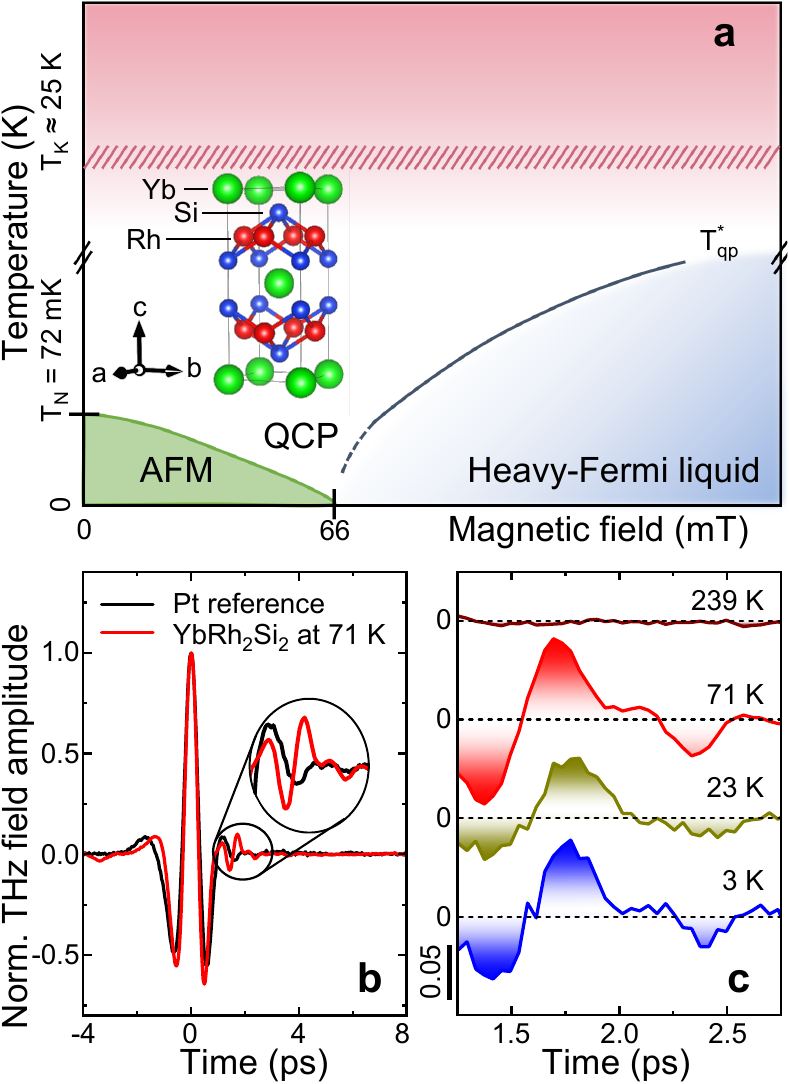}
\caption{\textbf{Exploring fermionic quantum criticality by time-resolved THz reflectivity.} \textbf{a,} Schematic phase diagram of YbRh$_{2}$Si$_{2}$ with characteristic energy scales. \textbf{b,} Reflected THz signal from the YbRh$_{2}$Si$_{2}$ sample (red) and from a Pt mirror reference (black). The time traces are normalized by the maximum field amplitude at $t=0$\,ps. The delayed, purely Kondo-related ``echo'' pulse is visible in the interval between 1.3 and 2.6\,ps, where it distinctly differs from the Pt reference signal. \textbf{c,} Signal of the delayed pulse in YbRh$_{2}$Si$_{2}$ at $B_{\perp}=0$ for various temperatures.}
\label{fig:YRS_1}
\end{figure}

At a continuous phase transition, the ordered and the disordered phases become equal in energy. As a consequence, the fluctuations between these two states become infinitely slow. This so-called critical slowing down (CSD) is universally observed in the dynamics of classical fields which are bosonic in nature and vanish at the phase transition, like the magnetization, associated with bosonic magnons, in the case of ferromagnetic order\cite{StatPhys}. By contrast, CSD of fermionic excitations or quasiparticles is generally not expected to occur since fermions as elementary particles are thought to be indestructible. However, certain quantum materials known as heavy-fermion (HF) compounds host \textit{composite} fermionic quasiparticles. These are quantum superpositions of itinerant and localized, i.e., heavy, electron states generated by the Kondo effect\cite{Kondo1964,Hewson1993}, with a low binding energy parameterized by the Kondo energy scale, the lattice Kondo temperature $T_K$. At a magnetic quantum phase transition (QPT) in such materials (see Fig.~\ref{fig:YRS_1}a), these brittle, heavy quasiparticles are assumed to disintegrate due to critical fluctuations\cite{Si2001,Coleman2001,Loehneysen2007} despite their fermionic nature, and fermionic CSD would be a unique signature of this destruction. 

Using time-resolved terahertz (THz) spectroscopy, we directly observe such fermionic CSD as a suppression of the heavy-particle hybridization gap and flattening of the associated band. This ``softening'' expands the region in momentum space where resonant THz absorption is allowed. We observe this as an \textit{increase} instead of a Kondo-weight-loss-generated decrease of the Kondo-related THz signal towards the quantum critical point (QCP) before the heavy quasiparticle band vanishes altogether below a breakdown temperature $T_{\rm qp}^*$\cite{Gegenwart2007}. Moreover, we identify a critical exponent in this behaviour and thus set the stage for the classification of fermionic quantum criticality in analogy to the criticality of thermodynamic phase transitions. 

Signatures of Kondo quasiparticle destruction were suspected from the dynamical scaling of the magnetic susceptibility\cite{Schroeder2000}, specific heat measurements\cite{Custers2003}, Hall effect measurements of the carrier density\cite{Paschen2004}, and optical conductivity measurements\cite{Prochaska2020}. The conjectures entering the interpretation of these measurements have been challenged, however\cite{Vojta2008,Woelfle2011}. Time-resolved THz spectroscopy is a unique tool to probe the heavy quasiparticle dynamics and to resolve these questions. Specifically, HF materials respond to an incident ultrashort THz pulse by the emission of a time-delayed reflex\cite{Wetli2018}. This ``echo'' is a response from the reconstructing Kondo ground state after its destruction by the incident pulse. Hence, time acts as a filter separating the Kondo-sensitive delayed pulse from the Kondo-insensitive main pulse so that the former is a \textit{background-free response of the HF state}. Specifically, the amplitude and the delay time of the echo pulse are proportional to the HF spectral weight and the Kondo coherence time $\tau_K=2\pi\hslash/k_BT_K$, respectively (with $\hslash$ the Planck and $k_B$ the Boltzmann constant). As a method, this technique was introduced through experiments on CeCu$_{6-x}$Au$_{x}$\cite{Wetli2018,Pal2019,Yang2020}.

In YbRh$_2$Si$_2$ we now apply this new method to directly measure the fermionic CSD. YbRh$_2$Si$_2$ is a prototypical HF compound. In zero magnetic field it is antiferromagnetic below the Ne\'el temperature $T_N=72$\,mK. It undergoes a QPT to a Kondo HF liquid at a critical magnetic field of $B_\perp \approx 66$\,mT perpendicular to the $c$ axis\cite{Trovarelli2000,Gegenwart2002} (see Fig.~\ref{fig:YRS_1}a)) induced by the  Ruderman-Kittel-Kasuya-Yosida (RKKY) magnetic interaction between the Yb  moments\cite{Ruderman1954,Kasuya1956,Yosida1957,Nejati2017}. Alternatively, a 6\% substitution of Rh by Ir creates a QCP at zero field\cite{Krellner2012,Friedemann2009}. YbRh$_2$Si$_2$ has a Kondo temperature of $T_K\approx 25$~K, high enough to enable a wide quantum-critical region and to permit us to search for signs of CSD in the range $T^*_{\rm qp}<T<T_K$, in contrast to the case of CeCu$_{6-x}$Au$_{x}$. In our temperature-dependent, time-resolved THz reflection spectroscopy measurements we crossed the QCP by varying the magnetic field or the Ir concentration. The 1.5-cycle THz pulses of $\sim 2$\,ps duration were incident onto the $c$-cut Yb(Rh$_{1-x}$Ir$_{x}$)$_2$Si$_2$ samples ($x=$0, 0.06). The echo pulses were analysed as described elsewhere\cite{Wetli2018}. With $T_K=25$\,K we obtain a delay time  $\tau_K\approx 1.9$\,ps for these in agreement with the data in Figs.~\ref{fig:YRS_1}b and \ref{fig:YRS_1}c. We therefore chose the time window for the analysis from $1.3$ to $2.6$\,ps.

\begin{figure*}
\centering
\includegraphics[width=1.9\columnwidth]{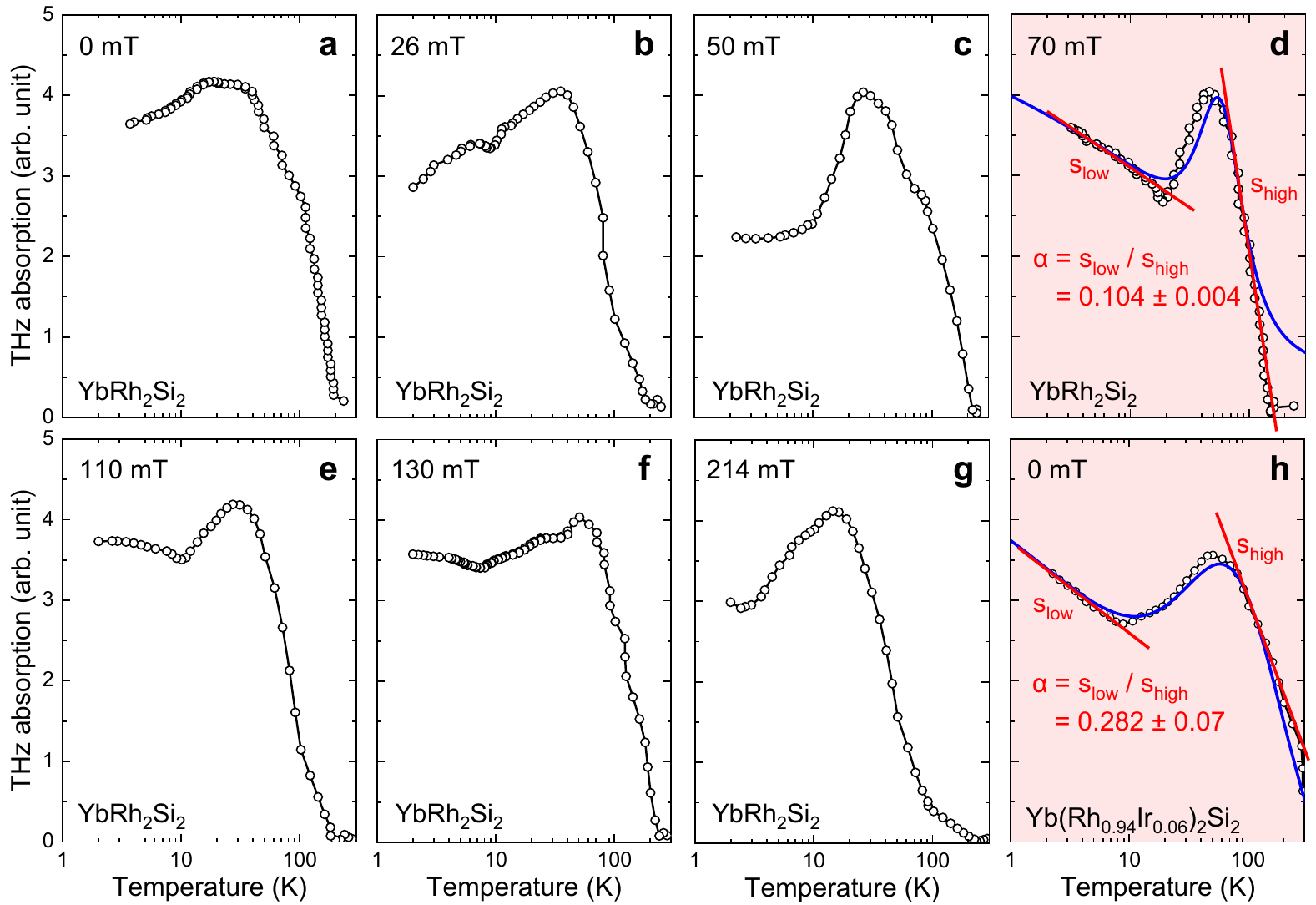}
\caption{\textbf{Temperature dependence of the resonant THz absorption across the QCP in YbRh$_{2}$Si$_{2}$.} \textbf{a}--\textbf{g,} Evolution of the THz absorption by resonant Kondo quasiparticle excitations at various external magnetic fields $B_{\perp}$ as a function of temperature. The weights are derived from the integrated intensity of the echo pulses emitted in the time window of $1.3$ -- $2.6$\,ps time window (see Figs.~\ref{fig:YRS_1}b, c). \textbf{h,} Temperature dependence of the resonant THz absorption in quantum critical Yb(Rh$_{1.94}$Ir$_{0.06}$)$_2$Si$_2$. Fits of Eq.~\ref{eq:THz_absorption} are plotted as blue lines in \textbf{d} and \textbf{h}.}
\label{fig:YRS_2}
\end{figure*}

Fig.~\ref{fig:YRS_2} shows the time-integrated intensity of the THz echo pulse for Yb(Rh$_{1-x}$Ir$_{x}$)$_2$Si$_2$ for $x=0$ and $B_\perp=214$\,mT and for $x=0.06$ and $B_\perp=0$. All plots exhibit a similar maximum value of the spectral Kondo weight near $25$\,K in good agreement with $T_K$. Aside from a shift towards higher temperature right at the QCP, the maximum shows no systematic field dependence. Below the peak temperature, the signal initially decreases with temperature for all magnetic fields. This can be attributed to the reduced thermal broadening of THz-induced interband transitions and is reproduced by the theory introduced below.

On the antiferromagnetic side of the QCP (Figs.~\ref{fig:YRS_2}a and \ref{fig:YRS_2}b), the signal continues to decrease but remains finite down to the lowest experimentally achieved temperature of 2\,K. We note that this temperature and field range (2.0\,K~$\leq T\leq$~20\,K, $B_\perp\lesssim 66$\,mT) is within the white area of the phase diagram in Fig.~\ref{fig:YRS_1}a, the so-called quantum-critical fan\cite{Custers2003,Friedemann2009}, where thermodynamic and transport properties are dominated by quantum-critical fluctuations\cite{Trovarelli2000,Gegenwart2002,Prochaska2020}. 
However, our THz time-delay spectroscopy is not directly sensitive to these fluctuations, but exclusively to the HF quasiparticle spectral weight\cite{Wetli2018,Pal2019}. Therefore, the behavior in  Figs.~\ref{fig:YRS_2}a and \ref{fig:YRS_2}b indicates that the Kondo effect remains partially intact in this temperature range. This is reasonable since we are still a factor of $\sim 25$ above $T_N$ so that the heavy quasiparticles are not entirely destroyed by the impending antiferromagnetic order..

At quantum criticality (Figs.~\ref{fig:YRS_2}d and \ref{fig:YRS_2}h), the temperature dependence changes drastically. The initial signal decrease with temperature is now followed by a logarithmic \textit{increase} of the THz echo signal that persists down to the lowest observed temperature. Note qualitative similarity between the field-tuned (Fig.~\ref{fig:YRS_2}d) and chemically tuned (Fig.~\ref{fig:YRS_2}h) quantum critical systems, however with different logarithmic slopes. On the HF liquid side of the QCP the signal increase towards the lowest temperatures is still present and gradually fades away with distance from the QCP (Figs.~\ref{fig:YRS_2}e-g). Its onset may also be conjectured  at $50$~mT on the antiferromagnetic side. (Fig.~\ref{fig:YRS_2}c).   

\begin{figure*}
\centering
\includegraphics[width=1.9\columnwidth]{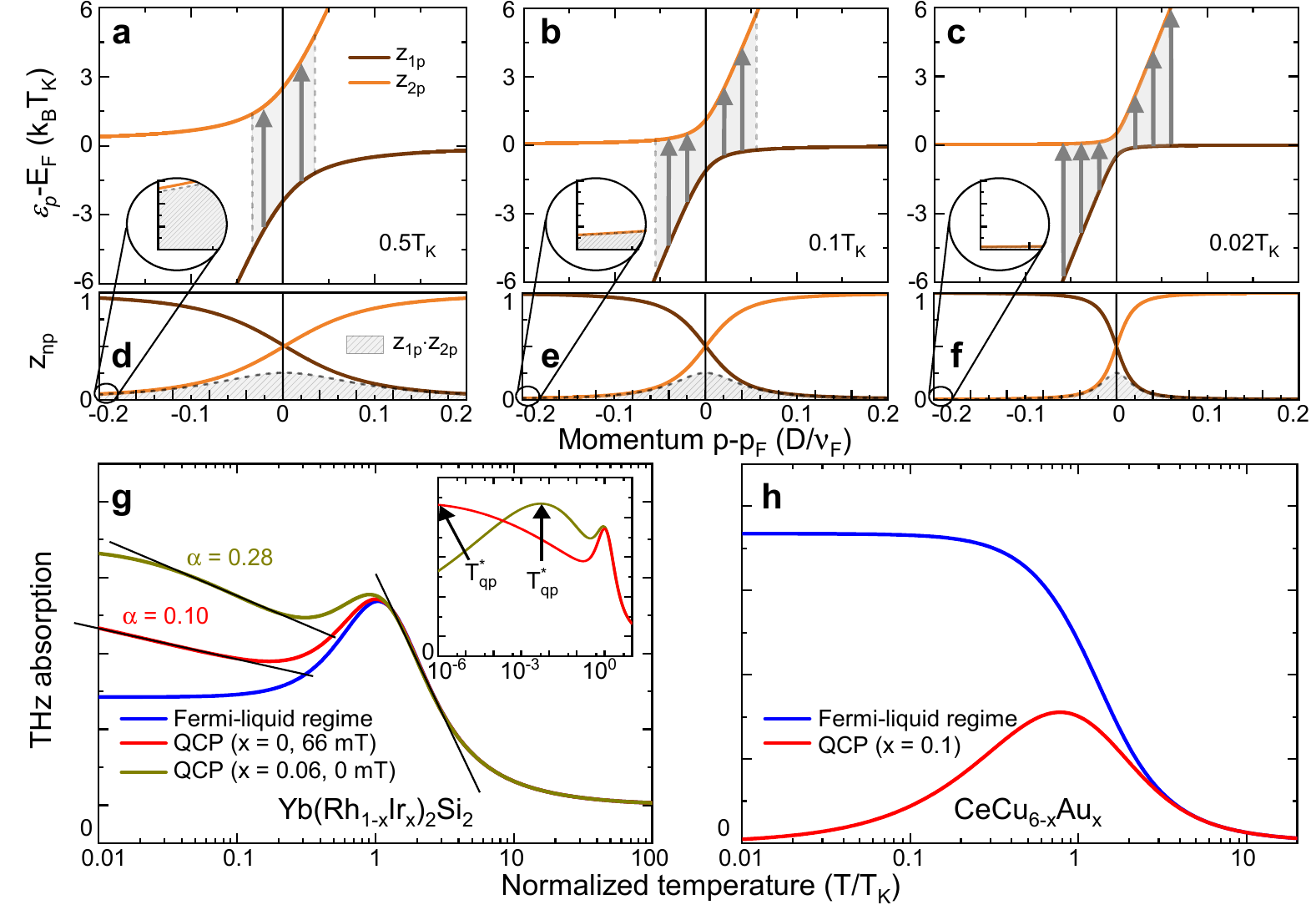}
\caption{\textbf{Band structure and Kondo weight calculations towards the QCP.} \textbf{a}--\textbf{c,} Band structure of the  conduction band (steep slope) and the Kondo state (flat band) resulting in a hybridized lower (brown) and upper (orange) branch. \textbf{d}--\textbf{f,} Momentum- and 
temperature-dependent quasiparticle weights $z_{1\vec{p}}$, $z_{2\vec{p}}$ in the lower (brown) and upper (orange) bands as well as the product $z_{1\vec{p}}\cdot z_{2\vec{p}}$, all calculated from the two-band critical Fermi liquid theory. \textbf{g}, \textbf{h,} Resonant THz absorption strength at (red and dark yellow) and away from (blue) the QCP as calculated for the system parameters of Yb(Rh$_{1-x}$Ir$_x$)$_{2}$Si$_2$, ($x=0,\,0.06$) and CeCu$_{6-x}$Au$_x$, respectively.} 
\label{fig:YRS_3}
\end{figure*}

To understand the striking logarithmic increase towards low temperature, we analyze the THz echo signal theoretically. In a HF system, the strongly correlated, flat band produced by the Kondo effect hybridizes with the light conduction band to generate a structure with a lower $(n=1)$ and an upper $(n=2)$ band with avoided crossing\cite{Paschen2016,Feng2017,Pal2019}, as shown in Figs.~\ref{fig:YRS_3}a to \ref{fig:YRS_3}c. While low-temperature thermodynamic and transport experiments probe the lower, occupied band only, resonant THz spectroscopy involves transitions between both bands, which calls for a two-band theory. Our according, critical HF liquid theory shows (see Methods) that the $n=1,2$ bands with dispersions $\varepsilon_{n\vec{p}}$ (where $\vec{p}$ is the crystal-electron momentum) have distinct momentum- and temperature-dependent spectral weights $z_{n\vec{p}}$. These cross over from $z_{n\vec{p}}\approx 1$ in the strongly dispersive region to $z_{n\vec{p}}=a\,z_0\ll 1$ in the flat region of these bands, see Figs.~\ref{fig:YRS_3}d to \ref{fig:YRS_3}f and insets. Here, $a\ll 1$ is the spectral weight of the local, single-ion Kondo resonance, which builds up logarithmically from above $T_K$ and then saturates towards a constant value for $T<T_K$. Further, $z_0=(T/T_0)^\alpha$ is a suppression factor with a critical exponent $\alpha$. It describes the destruction of quasiparticle spectral weight as the QCP is approached by lowering the temperature $T$ below the onset temperature for quantum criticality, $T_0$. For the latter, experiments revealed $T_0\approx T_K$ in YbRh$_2$Si$_2$\cite{Gegenwart2006}.

As mentioned, the THz echo pulse at $\tau_K=2\pi\hslash/k_BT_K$ is solely sensitive to the breakup-and-recovery dynamics of the HFs and not to other THz absorption channels\cite{Wetli2018,Pal2019,Yang2020}. Therefore, its intensity exclusively depends on the quasiparticle weight and the phase space available for THz-induced excitations. Specifically, the echo-pulse intensity is proportional to the probability
\begin{equation} \label{eq:THz_absorption}
  P(T)= A \int d^3p\ z_{1\vec{p}}\ z_{2\vec{p}}\ f(\varepsilon_{1\vec{p}})\,[1-f(\varepsilon_{2\vec{p}})]\ W(\Delta \varepsilon_{\vec{p}})
\end{equation}
for the resonant excitation of electrons from the lower to the upper band at an energy difference $\Delta\varepsilon_{\vec{p}}=\varepsilon_{2\vec{p}}-\varepsilon_{1\vec{p}}$. Here, $f(\varepsilon_{n\,\vec{p}})$ is the Fermi-Dirac distribution function. In Eq.~(\ref{eq:THz_absorption}), it describes the probability that the $n=1$ band is occupied and the $n=2$ band is empty before the THz absorption process. $W(\hslash\omega)$ is the spectrum of the incident THz pulse, which is a Gaussian distribution of width $\Gamma$ centered around the central frequency $\Omega_{\rm THz}$. With $\Delta\varepsilon_{\vec{p}}\approx \hslash\Omega_{\rm THz}$, the THz-induced interband transition becomes resonantly allowed. The integral runs over all electron momenta, and the factor $A$ is a temperature-independent constant, proportional to the intensity of the incident THz pulse and to the modulus square of the electric-dipole transition-matrix element between the two bands.

When the probability for HF formation, $a\,z_0\propto (T/T_K)^{\alpha}$, tends to zero at the QCP, the heavy bands flatten according to the two-band HF liquid theory. Also, the hybridization gap vanishes, and with that the quasiparticle energy in the heavy regions of both bands ($n=1,2$) approaches the Fermi energy $E_F$, see Figs.~\ref{fig:YRS_3}d to \ref{fig:YRS_3}f. This means that the oscillation frequency of fermionic quasiparticles, $\omega_{n\vec{p}}=(\varepsilon_{n\vec{p}}-E_F)/\hslash$, vanishes, which is indicative of a fermionic CSD. In turn, it implies an expansion of the region in the momentum space where resonant THz transitions are allowed, seen as broadening of the shaded areas in Figs.~\ref{fig:YRS_3}a to \ref{fig:YRS_3}c. The interplay of these two counteracting effects, quasiparticle destruction and phase-space expansion, leads to a non-monotonic temperature dependence of the THz absorption strength $P(T)$. An expansion of Eq.~(\ref{eq:THz_absorption})for small $z_0(T)$ predicts a logarithmic \textit{increase} of $P(T)\propto \ln [1/z_0(T)]$ towards low temperatures down to the region of $T_{\rm qp}^*$. The full numerical evaluation of Eq.~\ref{eq:THz_absorption} leads to the behavior shown in Fig.~\ref{fig:YRS_3}g. It reproduces the Kondo maximum near $T_K\approx 25$\,K and at quantum criticality indeed to a logarithmic increase across an intermediate temperature window $T_{\rm qp}^*<T<T_K$ according to  
\begin{equation} \label{eq:log}
  P(T)=\alpha \,A\, \ln (T_K/T)\qquad,
\end{equation}
observing this behavior in Figs.~\ref{fig:YRS_2}d and \ref{fig:YRS_2}h is thus a unique experimental signature of fermionic quasiparticle CSD in the Yb(Rh$_{1-x}$Ir$_{x}$)$_2$Si$_2$ system. 

Upon further decreasing the temperature to $T<T^*_{\rm qp}$, $P(T)$ approaches zero as the HF weight disappears altogether, see inset of Fig.~\ref{fig:YRS_3}g. The low-temperature scale $T_{\rm qp}^*$ is thus defined as the position of the signal maximum between the logarithmic increase and its ultimate collapse towards $T\to 0$. As seen in Fig.~\ref{fig:YRS_3}g (inset), $T_{\rm qp}^*$ depends on the critical exponent $\alpha$ and is several orders of magnitude lower than the Kondo scale $\sim T_K$, possibly undetectably small. A low-temperature scale $T^*$ has also been observed as a maximum in the magnetic susceptibility\cite{Friedemann2009} whose microscopic origin has, however, remained unclear. Since in Yb(Rh$_{1-x}$Ir$_{x}$)$_2$Si$_2$, $T^*$ and our theoretically predicted $T_{\rm qp}^*$ are in the same temperature range, we conjecture that both may be of the same physical origin - the competition of fermionic CSD and quasiparticle breakdown at the QCP. As a crossover temperature, $T_{\rm qp}^*$ remains non-zero, but can be exceedingly small, depending on $\alpha$, see inset of Fig.~\ref{fig:YRS_3}g. Note that away from criticality, we have $\alpha\to 0$ so that the logarithmic low-temperature behavior does not occur in agreement with Figs.~\ref{fig:YRS_2}e to \ref{fig:YRS_2}g. The blue curves in Figs.~\ref{fig:YRS_2}d and \ref{fig:YRS_2}h represent the evaluation of Eq.~(\ref{eq:THz_absorption}) for the spectrum $W(\hslash\omega)$ of the THz pulses used in our experiment. Considering that $\alpha$ is the only adjustable parameter apart from the overall signal amplitude and that we use the same value $T_K\approx 25$\,K for both curves, the agreement between theory and data is excellent.

We can now extract the critical exponent $\alpha$ by comparing the logarithmic slope $s_{\mathrm{low}}$ associated with the the CSD at low temperatures ($T^*_{\rm qp}<T<T_K$) from Eq.~(\ref{eq:log}) with the slope $s_{\mathrm{high}}$ of the standard logarithmic behaviour of the Kondo weight at high temperature ($T>T_K$) according to $P(T)=A\,\ln(T_K/T)$. This directly leads to $\alpha=s_{\mathrm{low}}/s_{\mathrm{high}}$. From the experimental data we find $\alpha=0.10$ for Yb(Rh$_{1-x}$Ir$_{x}$)$_2$Si$_2$ at $x=0$, $B_\perp=66$\,mT in Fig.~\ref{fig:YRS_2}d and $\alpha=0.28$ at $x=0.06$, $B_\perp=0$ in Fig.~\ref{fig:YRS_2}h. Note that different critical behavior for QPTs by magnetic fields and chemical pressure have been observed before\cite{Loehneysen2001}. It is a signature of the different types of quantum-critical fluctuations in the two cases. 

Our theory also explains why the logarithmic low-temperature increase of THz absorption indicating CSD cannot be observed in the CeCu$_{6-x}$Au$_x$ system. For this material, the ratio $T_K/T^*$ is substantially smaller than in the Yb(Rh$_{1-x}$Ir$_{x}$)$_2$Si$_2$ system so that the effects of the buildup of Kondo weight and of the CSD overlap to such an extent that the latter is obscured, as seen in Fig.~\ref{fig:YRS_3}h and in agreement with experiment\cite{Wetli2018}.

\begin{figure}
\centering
\includegraphics[width=\columnwidth]{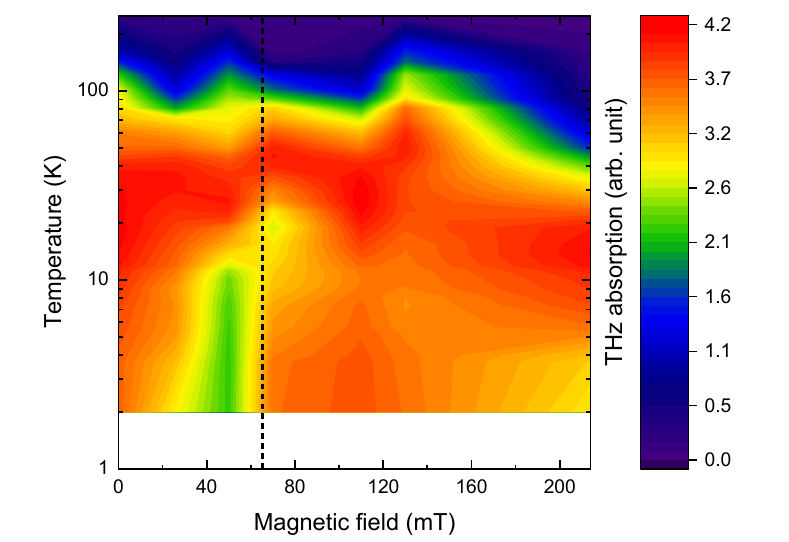}
\caption{\textbf{Phase diagram of field-tuned YbRh$_{\mathbf{2}}$Si$_{\mathbf{2}}$} as measured by THz time-delay spectroscopy, using the discrete magnetic-field values shown in Fig.~\ref{fig:YRS_2}.  The color code represents the resonant THz absorption, and the dashed line marks the critical field of 66\,mT. }
\label{fig:YRS_4}
\end{figure}

Summarizing in Fig.~\ref{fig:YRS_4} the measured THz absorption data in a $T$-versus-$B$ phase diagram reveals that on the HF liquid side of the QCP ($B_\perp > 66$\,mT) $P(T)$ is enhanced for $T<10$\,K due to the CSD effect and as explained by our two-band HF liquid theory. By contrast, on the antiferromagnetic side ($B_\perp \lesssim 50$\,mT) we observe a reduction of the Kondo-weight-related absorption but no fermionic CSD. This suggests that in this region the RKKY interaction strongly affects the quasiparticle dynamics such that the the two-band HF liquid theory is not valid here\cite{Ruderman1954,Kasuya1956,Yosida1957}. 

To conclude, we observed a logarithmic low-temperature increase in the resonant quasiparticle excitation probability $P(T)$ near a magnetic quantum phase transition in heavy-fermion materials. We identified this logarithmic increase as a unique signature of fermionic quasiparticle CSD, that is, a vanishing quasiparticle frequency, near a quantum phase transition with fermionic breakdown. Since, in contrast to thermodynamic and transport properties, our time-resolved THz spectroscopy is exclusively sensitive to the HF quasiparticle dynamics as opposed to thermal fluctuations, we could further extract the fermionic critical exponent $\alpha$ of the vanishing quasiparticle weight. The critical behaviour of $\alpha$ suggests to define the heavy quasiparticle weight as an order parameter for quantum phase transitions with fermionic breakdown. It also sets the stage for a classification of fermionic quantum phase transitions in terms of their critical exponent, analogous to thermodynamic phase transitions. \\

\noindent{\bf Methods}\\
\noindent{\bf Experimental}\\
Single-crystalline, $c$-oriented Yb(Rh$_{1-x}$Ir$_{x}$)$_2$Si$_2$ platelets ($x=0,\,0.06$) with dimensions of $2\times 3\times 0.07$\,mm$^3$ were grown from indium flux as described in the literature\cite{Krellner2012}. The sample surface is freshly polished before the THz measurements. Samples are mounted onto a Teflon holder, where two permanent magnets placed above and below the sample generate a magnetic field of up to 214\,mT 
in the easy magnetic plane perpendicular to the tetragonal \textit{c}-axis. We use a temperature-controlled Janis SVT-400 helium-reservoir cryostat operable in the range from 1.9 to 325\,K.  The THz experiments are performed in a $90^\circ$ reflection geometry with THz light in a spectral range from 0.1 to 3\,THz polarized perpendicular to the crystallographic \textit{c}-axis.

We generate single-cycle THz pulses by optical rectification in a 0.5-mm-thick (110)-cut ZnTe single crystal, using 90\% of an amplified Ti:Sapphire laser output (wavelength 800\,nm, pulse duration 50\,fs, pulse repetition rate 1\,kHz, 2.5\,mJ pulse energy). The energy of the THz pulse is in the range of a few nJ. The residual 10\% of the 800-nm beam is then used for free-space electro-optic sampling of the reflected THz light from the sample. Both, the THz and the 800-nm beams, are collinearly focused onto a 0.5-mm-thick (110)-cut ZnTe detection crystal. To increase the accessible time delay between the THz and the 800-nm pulses, Fabry-P\'erot resonances from the faces of the detection crystal are suppressed by a 2-mm-thick THz-inactive (100)-cut ZnTe crystal. The THz-inactive crystal is optically bonded to the back of the detection crystal. The THz-induced ellipticity of the 800-nm beam is measured using a quarter-wave plate, a Wollaston prism and a balanced photodiode.\\

\noindent{\bf Theoretical}\\
We construct a phenomenological, critical two-band Fermi liquid theory to describe the THz-induced resonant transitions from the heavy conduction to the light valence band. Electrons in the light band have the dispersion $\varepsilon_p^{(0)}$ measured relative to the Fermi energy $E_F$, have spectral weight unity, and are assumed to be non-interacting. The parameters of the heavy electron states, generated by the Kondo effect, are controlled by the Kondo scale $T_K$. Their energy lies close to the Fermi level $E_F$, shifted by $\Delta\approx \pm k_B T_K$ above (particle-like HF systems) or below (hole-like HF systems) $E_F$\cite{Hewson1993,Feng2017}. They have a strongly reduced spectral weight $a(T)$ which reaches $a(0)= T_K/\gamma\ll 1$ at $T=0$ and decreases logarithmically for temperatures $T>T_K$\cite{Hewson1993}. Here, $\gamma$ is the effective hybridization of the rare-earth $4f$ orbitals with the conduction electron states. We also take a residual interaction of the quasiparticles within the heavy band into account. It implies an additional reduction of both, the quasiparticle weight and of the heavy band shift, by the local quasiparticle weight factor $z_0(T)$ (see main text), $a \to z_0\,a$, $\Delta\to z_0\Delta$. Taking now the hybridization $V$ between the light, uncorrelated band and the heavy band into account and calculating the hybridized band structure, the band energies are obtained as 
\begin{equation}
\varepsilon_{1,2\,\vec{p}} =  \frac{1}{2}\left[
  \varepsilon_{\vec{p}}^{(0)} + z_0\Delta \pm \sqrt{(\varepsilon_{\vec{p}}^{(0)}-z_0\Delta)^2
  + 4z_0a|V|^2}
    \right],  
\end{equation}
and are shown, for different temperatures, in Figs.~\ref{fig:YRS_2}d to  ~\ref{fig:YRS_2}f. Due to the hybridization, the quasiparticle weights in the lower (1) and upper (2) bands become momentum dependent and read 
\begin{equation}
  z_{1,2\,\vec{p}} = \frac{(1+z_0a)(\varepsilon_{1,2\,\vec{p}}-z_0\Delta)- z_0a (\varepsilon_{\vec{p}}^{(0)}-z_0\Delta)}{2(\varepsilon_{1,2\,\vec{p}}-z_0\Delta)- (\varepsilon_{\vec{p}}^{(0)}-z_0\Delta)}\ ,
\end{equation}
as shown in Figs.~\ref{fig:YRS_2}a to ~\ref{fig:YRS_2}c. Inserting these expressions into Eq.~(\ref{eq:THz_absorption}) and adjusting the two parameters $T_{\rm K}/D$ and $\alpha$, where $D$ is the free conduction bandwidth, leads to the curves shown in Figs.~\ref{fig:YRS_2}d, h and \ref{fig:YRS_3}g, h. Symmetry implies that this result for the THz absorption is the same for particle-like and hole-like HF systems. \\

\noindent{\bf Acknowledgement.}\\
This work was financially supported by the Swiss National Science Foundation (SNSF) via project No. 200021\_178825 (M.F., S.P., C.J.Y.) and by the Deutsche Forschungsgemeinschaft (DFG) via SFB/TR 
185 (277625399) OSCAR and the Cluster of Excellence ML4Q (90534769, project C4) (J.K.) as well as via SFB/TRR 288 (422213477, project A03) and grant no. KR3831/4-1 (K.K, C.K.). S.P. further acknowledges the support by ETH Career Seed Grant SEED-17 18-1.\\

\noindent{\bf Author contributions.}\\
All authors contributed to the discussion and interpretation of the experiment and to the completion of the manuscript. S.P. and C.J.Y. performed the experiment and the data analysis. K.K. and C.K. provided the YbRh$_2$Si$_2$ and Yb(Rh$_{0.94}$Ir$_{0.06}$)$_2$Si$_2$ samples. J.K. performed the theoretical analysis. J.K. and M.F. conceived the work, and S.P. supervised the experiments. S.P., J.K. and M.F. drafted the manuscript.\\

\noindent{\bf Competing interests.}\\
The authors declare that they have no competing financial interests.\\

\noindent{\bf Correspondence.}\\
Correspondence and requests for materials should be addressed to, M.F. (email: manfred.fiebig@mat.ethz.ch), J.K. (email: kroha@physik.uni-bonn.de) or S.P. (email: shovon.pal@niser.ac.in)

\end{document}